\documentstyle[12pt,psfig]{article}

\newcommand{\AmS}{{\protect\the\textfont2
\renewcommand{\thesection}{\Roman{section}}
  A\kern-.1667em\lower.5ex\hbox{M}\kern-.125emS}}
 
\begin{document}
\rightline {DFTUZ 98/11}
\vskip 2. truecm
\centerline{\bf Frustration in Finite Density QCD}
\vskip 2 truecm
\centerline { R. Aloisio$^{a,d}$, V.~Azcoiti$^b$, G. Di Carlo$^c$, 
A. Galante$^{a,c}$ and A.F. Grillo$^d$}
\vskip 1 truecm
\centerline {\it $^a$ Dipartimento di Fisica dell'Universit\`a 
dell'Aquila, L'Aquila 67100 (Italy).}
\vskip 0.15 truecm
\centerline {\it $^b$ Departamento de F\'\i sica Te\'orica, Facultad 
de Ciencias, Universidad de Zaragoza,}
\centerline {\it 50009 Zaragoza (Spain).}
\vskip 0.15 truecm
\centerline {\it $^c$ Istituto Nazionale di Fisica Nucleare, 
Laboratori Nazionali di Frascati,}
\centerline {\it P.O.B. 13 - Frascati 00044 (Italy). }
\vskip 0.15 truecm
\centerline {\it $^d$ Istituto Nazionale di Fisica Nucleare, 
Laboratori Nazionali del Gran Sasso,}
\centerline {\it Assergi (L'Aquila) 67010 (Italy). }
\vskip 3 truecm

\centerline {ABSTRACT}
\vskip 0.5truecm

\noindent
We present a detailed analysis of the QCD partition function in the
Grand Canonical formalism. Using the fugacity expansion we find 
evidence for numerical instabilities in the standard evaluation 
of its coefficients. 
We discuss the origin of this problem and propose an issue to it. 
The correct analysis shows no evidence for a discontinuity in the 
baryonic density in the strong coupling limit. The moderate optimism 
that was inspired by the Grand 
Canonical Partition Function calculations in the last years has to be 
considered ill-founded.

\vfill\eject

Non perturbative investigations of finite density QCD have received a 
growing attention in the last years. Concerning the most powerful non 
perturbative approach, numerical simulations on a lattice, 
there is some evidence that the long standing problem
of dealing with a complex valued determinant can be overcome using 
non standard numerical approaches based on the calculation of 
the Grand Canonical Partition Function (GCPF). 
This technique has attracted much attention since,
once evaluated the coefficients in the fugacity expansion, it
allows free mobility in the chemical potential $\mu$ at negligible computer 
cost.

The strong coupling limit is by far the most investigated since it is possible
to check numerical results with analytical predictions \cite{ANALY} and 
monomer-dimer
simulations \cite{KARSCH}. Using the Glasgow algorithm 
evidence for a 
first order phase transition at a value of the chemical potential in good
agreement with the old results of the monomer-dimer
simulations was found \cite{BARBOUR}.

In a previous paper \cite{NOI} 
we have performed simulations with a Microcanonical 
Fermion Average (MFA) \cite{MFA} 
inspired method, using the modulus of the fermionic 
determinant to define a real and positive definite effective action. 
Concerning the baryon density, we have found essential agreement with 
the results of \cite{BARBOUR}. 
In that work we have completely diagonalized the propagator matrix $P$ 
\cite{GIBBS} 
and the eigenvalues have been used to calculate the coefficients of the 
fugacity expansion through a standard recursion method.
We found evidence for a first order phase transition at zero and non
zero masses up to $m\sim 0.7$ and vanishing of such a signal for larger
masses. The critical chemical potential $\mu_c$
was in good agreement with the one of Karsch and M\"utter \cite{KARSCH} 
for small 
and intermediate masses. The lacking of the transition at larger masses 
is, on physical grounds, an unexpected result, previously reported also in 
\cite{BARBOUR}.

In addition to the diagonalization of the quark propagator matrix, we also 
performed in \cite{NOI} a direct diagonalization of the Dirac operator 
$(\Delta)$ in a $4^4$ lattice. 
At large values of $m$ the results were in contradiction. The latter method 
allowed a clear
determination of a first order transition with a critical chemical potential 
$\mu$ in agreement with an extrapolation of the data of \cite{KARSCH} 
and with quenched simulations \cite{MPL},  
whereas the former approach, based on the 
calculation of the coefficients in the fugacity expansion, 
did not produce any signal of first order transition. 

These contradictory results point to the existence of numerical 
problems in the evaluation of  
thermodynamical quantities. Three are the possible sources for numerical 
troubles: i) the evaluation
of the eigenvalues of the fermion matrix $(\Delta)$, ii) the 
diagonalization of the propagator matrix $(P)$ and iii) the determination
of the fugacity expansion coefficients.

We used a standard NAG library routine to perform the diagonalization
of $\Delta$ and $P$. At $\mu=0$, we found a perfect agreement  
between the $\Delta$ eigenvalues computed with this routine 
with the ones obtained using a standard Lanczos algorithm. 
Using the two sets of eigenvalues, we have also 
verified the relation \cite{GIBBS} 
\begin{equation}\label{detd}
\left(\frac{1}{2}\right)^{3V} e^{3V\mu}\det(P-e^{-\mu})=\det\Delta.
\end{equation}
in the whole range considered for the chemical potential $\mu$ and quark 
mass $m$. 
We conclude that the diagonalization procedure is stable for any value of 
the mass and the chemical potential and that 
the numerical problems
can only be due to the manipulations necessary to go from the $P$ 
eigenvalues to the Grand Canonical expansion coefficients $a_n$
\begin{equation}\label{detp}
e^{3V\mu}\det(P-e^{-\mu})=\sum_{n=-3V_s}^{3V_s}a_ne^{-nl_t\mu}.
\end{equation}
This conclusion was indeed corroborated by the numerical results at large 
fermion masses in $4^4$ and $6^3\times4$ 
lattices. The numerical results in this 
region of masses, when computed from the eigenvalues of $\Delta$ or from 
the eigenvalues of $P$, without making use of the fugacity coefficients, 
are in perfect agreement and show a clear signal for a first order phase 
transition, as discussed below. This signal was 
absent in the numerical 
results obtained from the fugacity coefficients.

\vskip 0.3truecm
\noindent
{\bf The single configuration analysis}
\vskip 0.3truecm

Having localized the source of numerical troubles in the determination 
of the fugacity coefficients and in order to get some insight of these 
numerical difficulties, we will first focus on a single random 
configuration of a $6^3\times4$ lattice. A property of the effective 
action which can be 
used to check the numerical calculations of the Grand Canonical expansion
coefficients is the parity under the transformation $\mu\to-\mu$
since, as well known, the real part of the determinant must be an 
even function of the chemical potential. 
In fig 1 we report the asymmetry  
$\log|Re(\det\Delta(\mu))|-\log|Re(\det\Delta(-\mu))|$ against the chemical 
potential $\mu$ at $m=0.1$. This quantity has been 
computed using the left and right hand sides of expression (\ref{detp})
with a 128 bit arithmetic.
It is evident that the coefficients do not respect the symmetry 
$a_{-n}=a_n^*$ while
the determinant, directly evaluated trough (\ref{detd}), behaves
as expected. 
These results hold in the chiral limit as well as for masses up to 1.5.

Again this analysis suggests the existence of perverse effects in the code 
used to determine the fugacity coefficients.
Therefore it is important to realize how these problems origin and how can 
be overcome to get full advantage of the Grand Canonical formalism.

To this end, and inspired by the strong coupling 
results \cite{KARSCH}, we have analysed a simple case. 
In the strong coupling limit we expect a first order saturation transition 
in the number density $n(\mu)$ which can be
approximated by a Heaviside $\theta$ function. 
To check our ideas we have considered a particular distribution of the 
$P$ eigenvalues that reproduces the number density with the expected behaviour.
In the infinite volume limit the radial distribution of the eigenvalues 
is related to the first derivative of the number density with respect to the 
chemical potential. 
To mimic a first order saturation transition separating a phase where 
$n(\mu)=0$ for 
$\mu<\mu_c$ from another phase where $n(\mu)=1$ for $\mu>\mu_c$ we have
considered the eigenvalue distribution 
$\lambda_i^\pm=e^{\pm\mu_c} e^{\pm i\theta_i}$, 
where the phases $\theta_i$ are uniformly distributed in $(-\pi,\pi)$.

The $a_n$ coefficients for this set of eigenvalues 
have been calculated 
with the same algorithm used for real data. This has been done either with
input eigenvalues ordered respect to their phases or randomly ordered.
In fig 2 we report the number density obtained in both cases.
It is evident that the former case leads to wrong results while the
latter reproduces the correct ones.

The origin of numerical instabilities in this model can be easily
understood. If we consider $2N$ eigenvalues uniformly distributed on two 
circles of radius $\rho$ and $\rho^{-1}$ the polynomial in the fugacity 
contains only three non vanishing terms: $a_{\pm N}=1$ and $a_0=\rho^N+\rho^{-N}$.  
If we calculate the coefficients with the standard recursion method we
use the first $n$ eigenvalues to calculate the coefficients of a polynomial
$P_n$ of degree $n$:
\begin{equation}\label{cn}
a_k^n = a_{k-1}^{n-1} - \lambda_n a_k^{n-1} \qquad\qquad k\ge 1
\end{equation}
where $a_k^{n}$ is the coefficient of order $k$ of $P_n$.
If the eigenvalues are phase-ordered, at any intermediate step we calculate
the coefficients of a polynomial whose zeros lay on an arc of circle of
increasing length.
These coefficients are non zero and of order $O(\rho^{n})$ . 
Once their logarithm is bigger than $q\log_{10}2$, where $q$ are the
bits of the mantissa in the floating point representation number
($q=113$ for $128$ bits arithmetic), 
rounding propagation prevents to obtain the correct answer.
This happens already for relatively small $N$ and forbids the symmetries
to be realized in the final results 
(see \cite{KAR} for a similar effect in a different context).
Randomly ordered eigenvalues modify this scenario since the symmetries are
(almost) enforced at each intermediate step as well as in the final
result. The coefficients of $P_n$ never grow too much and rounding
effects are better under control.

We have noticed that, for real simulations,  
the output of our diagonalization routine has part
of the eigenvalues with almost ordered phases. If we shuffle them before
starting the computation of the coefficients, we recover the 
$\mu$, $-\mu$ symmetry for each gauge configuration and the results 
are indistinguishable from the ones
obtained computing the determinant of the quark propagator matrix 
without making use of the coefficients. 
Therefore we expect that the anomalous behaviour observed in the naive 
Heaviside model 
is not peculiar of its (ad hoc) eigenvalue distribution but will remain valid  
also for the $P$ eigenvalues of actual simulations. 
A good way to cure the perverse effects, induced by the rounding in the 
routine which computes the fugacity coefficients, is to shuffle the full 
set of eigenvalues before starting the computation of the coefficients.

\vskip 0.3truecm
\noindent
{\bf Strong coupling results}
\vskip 0.3truecm

We have reanalyzed our $\beta=0$ data in $6^3\times4$ lattices with
different methods. 
The coefficients have been evaluated using for each 
configuration a random ordering of its eigenvalues. In order to define a real
and positive partition function we have considered the modulus of the 
fermionic determinant as in \cite{NOI}, or we set equal to zero the 
coefficients with negative real part \cite{BARBOUR} (Glasgow approach).
The results for the number density have been checked computing the same
quantity without the coefficients and are shown in figures 3 and 4.
 
Figure 3 contains the results for m=1.5, the value of the fermion mass at which 
contradictory results were found in \cite{NOI}. 
The diamonds stand for the results 
obtained through the use of the fugacity coefficients computed in the standard 
way ($6^3\times4$ lattice), to be compared with the results obtained from  
the coefficients computed from the randomized set of eigenvalues 
(solid line, $6^3\times4$ lattice), and with the ones corresponding to 
a direct diagonalization of the Dirac matrix $\Delta$ ($4^4$ lattice).
We see how, after introducing the randomization procedure in the computation 
of the fugacity coefficients, the contradiction reported in \cite{NOI} 
disappears 
and the numerical results for the baryon density, showing a clear first order 
phase transition, are in good agreement with those reported in \cite{KARSCH} 
and 
also with the quenched results \cite{MPL}. We want to remark also that if we 
do not take the absolute value of the fermion determinant and compute the 
coefficients of the Grand Canonical Partition Function as in \cite{BARBOUR} 
but using the randomization procedure, we get results indistinguishable 
from the solid line in Fig. 3. These are encouraging results since they  
solve a contradiction and agree with other reliable results and with 
physical expectations.

In figures 4-a and 4-b we plot the same quantity as in Fig. 3 but for $m=0.1$.
Fig. 4-a contains the results for the baryon density obtained using the 
logarithm of the 
absolute value of the fermion determinant as effective action \cite{NOI},  
whereas in Fig. 4-b the analysis has been done without taking the absolute 
value of the fermion determinant and setting to zero all negative 
averaged coefficients in 
the fugacity expansion \cite{BARBOUR}.  
In both figures the diamonds 
stand for the results obtained from the fugacity coefficients computed 
with the standard procedure. The solid lines in Figs. 4-a, 4-b correspond 
to the number density obtained from the fugacity coefficients computed 
using the randomized procedure previously described. Computing the modulus 
of the determinant  
directly from the eigenvalues of the quark 
propagator matrix $P$ (1), we obtain numerical results for the number density 
indistinguishable from the solid line in Fig. 4-a.

A very unpleasant consequence of the analysis reported in Fig. 4 
is that the numerical data, if correctly analysed, show no evidence for 
a first order phase transition, contrary to what previously reported 
\cite{BARBOUR}, 
\cite{NOI} (diamonds in Figs 4-a, 4-b). The perverse effects detected in the 
standard computation of the fugacity coefficients are relevant not only 
at large values of the fermion masses but also at m=0.1 and in the chiral 
limit. Preliminary results in a $8^3\times4$ lattice at $m=0$ confirm this 
scenario.

After we have removed all the possible numerical artifacts some comments are
necessary to evidenciate the physical meaning of the results.
The first observation concerns the distribution of the phase of the 
determinant. 
When we reach the onset $\mu$ and up to the saturation point
it becomes (within the errors) indistinguishable from a flat distribution
(in the $-\pi$, $\pi$ range).
The same flat distribution is observed for all the coefficients except for
the first, the last and the central one that are constrained to be real from
the eigenvalues symmetries.
With available statistics we have found no correlation between the phase 
and the modulus of the 
determinant. The consequence is that the averaged determinant 
is no longer a real and positive quantity as it should be.
The same happens for each averaged coefficient whose sign is completely
indeterminate. This can be easly checked splitting the configurations
set in two or more subsets and calculating independent averages:
the phases of different averages have again a flat distribution while 
the modulus converge reasonably well.
We can conclude that the averaged determinant and coefficients have not
converged and we can not rely on the results for finite density QCD. This can
explain the discrepancies between the data obtained from the Gran Canonical
Partition Function and the Monomer-Dymer-Polymer algorithm \cite{KARSCH}.

A second comment is in order, to explain why the Glasgow approach 
\cite{BARBOUR} and the modulus of the determinant give the same results.
To address this point it is important to note that the partition function of
finite density $SU(3)$ can be written as \cite{NOI}
\begin{equation}
{\cal Z} = {\cal Z}_{||} \; \langle e^{i\phi_\Delta}\rangle_{||}
\end{equation}
where ${\cal Z}_{||}$ is defined using the modulus of the determinant,
$\phi_\Delta$ is the phase of the determinant and $\langle\;\rangle_{||}$ 
is the average defined using $|\det \Delta|$ as weight.
The same formula is valid for the coefficients (that are partition function
at fixed number density) once we substitute $\det \Delta$ with $a_n$ and
$\phi_{a_n}$ to $\phi_\Delta$.
The difference between ${\cal Z}$ and ${\cal Z}_{||}$ (or between 
$\langle a_n\rangle$ and $\langle |a_n|\rangle$) is significant only if the
contribution of the phase is a quantity proportional to $e^{-V}$ \cite{NOI}.
Such contribution (if present) can be appreciated only if we reduce the 
statistical error to $O(e^{-V})$ and to this aim we need a number of 
indipendent measurements of the order of the exponential of lattice
volume.
The data of phase distributions suggest that this is the scenario for
finite density QCD. Without $O(e^V)$ configurations
the results will never be significantly different from the 
ones obtained using the modulus of the determinant. 
A recent simulation of a $2^4$ lattice in the strong coupling limit
shows clearly how $O(10^6)$ configurations are needed to reproduce the 
expected mean field results \cite{BIELEFELD}.
We can use the Glasgow approach or whatever else but the result will
never be different from ${\cal Z}_{||}$ with reasonable statistics on
larger lattices.

This can also clarify the behaviour of the onset $\mu$ observed using the
GCPF formalism: it goes to zero when $m\to 0$ in a way that seems to be
consistent with one half the pion mass \cite{BARBOUR}, \cite{NOI}. 
This is what we expect from the theory defined through the modulus of 
the determinant where baryons of vanishing mass should be present in the 
chiral limit \cite{DAVIES}.

The main conclusion which follows from this work is rather pessimistic and 
frustrating. 
Neither the Glasgow method \cite{BARBOUR} nor our absolute value of the 
fermion determinant based approach \cite{NOI} are able to reproduce the 
reliable results of Karsch and M\"utter in the strong coupling limit 
\cite{KARSCH}. The moderate optimism that was inspired by the Grand 
Canonical Partition Function calculations in the last years has to be 
considered ill-founded.

Preliminary analysis at larger values of $\beta$ give  
indications that the infinite gauge coupling limit is the worst situation.  
In fact many of the coefficients in the fugacity expansion, 
which are negative with finite statistics 
at $\beta=0$, become positive at larger values of $\beta$, as they should be. 
There is still some hope that these approaches work better in the 
physically interesting region of larger $\beta$ and it is therefore 
worthwhile to check if this is the case.

\vskip 0.2 truecm
This work has been partly supported through a CICYT (Spain) - INFN (Italy)
collaboration.

\newpage
\centerline{\bf Figure Captions}
\vskip 1 truecm
\begin{itemize}

\item{Fig. 1: Difference between $log|\Delta(\mu)|$ and $log|\Delta(-\mu)|$
vs $\mu$ for a single gauge configuration 
obtained using the coefficients calculated in the standard way (a),
and from the eigenvalues of the propagator matrix (b).}

\item{Fig. 2: Number density in the naive Heaviside model computed with
phase ordered eigenvalues (diamonds) and with shuffled ones (solid line).}

\item{Fig. 3: Number density in a $6^3\times4$ lattice at $\beta=0$ and 
$m=1.5$ obtained using the coefficients calculated in the 
standard way (diamonds) and with the shuffled eigenvalues (solid line);
the same quantity for a $4^4$ lattice, computed from the eigenvalues of
$\Delta$ (squares).}

\item{Fig. 4: Number density in a $6^3\times4$ lattice at $\beta=0$ and
$m=0.1$ obtained using the coefficients computed in the standard way 
(diamonds) and with the shuffled eigenvalues (solid line); 
fermionic effective action as in the Glasgow method (a) and from the
modulus of the fermion determinant (b).}

\end{itemize}
\end{document}